# Terahertz spectroscopy of crystal-field transitions in magnetoelectric TmAl$_3$(BO$_3$)$_4$

A. M. Kuzmenko[1], A. A. Mukhin[1], V. Vu. Ivanov[1], G. A. Komandin[1], A. Shuvaev[2], A. Pimenov[2], V. Dziom[2], L. N. Bezmaternykh[3], I. A. Gudim[3]

[1]*Prokhorov General Physics Institute, Russian Academy of Sciences, 119991 Moscow, Russia*
[2]*Institute of Solid State Physics, Vienna University of Technology, 1040 Vienna, Austria*
[3]*L.V. Kirensky Institute of Physics, Siberian Branch of RAS, 660036 Krasnoyarsk, Russia*

**Abstract**

Dynamic magnetic properties of magnetoelectric TmAl$_3$(BO$_3$)$_4$ borate have been investigated by terahertz spectroscopy. Crystal field (CF) transitions within the ground multiplet $^3H_6$ of Tm$^{3+}$ ions are observed and they are identified as magnetic-dipole transitions from the ground singlet A$_1$ to the next excited doublet E of Tm$^{3+}$ ions. Unexpected fine structure of the transitions is detected at low temperatures. The new modes are assigned to local distortions of the sites with D$_3$ symmetry by Bi$^{3+}$ impurities, which resulted in the splitting of A$_1\rightarrow$E transition. Two types of locally distorted sites are identified and investigated. The main contribution to the static magnetic susceptibility is shown to be determined by the matrix elements of the observed magnetic transitions. We demonstrate that even in case of local distortions the symmetry of the undistorted crystal is recovered for magnetic and for quadratic magnetoelectric susceptibilities.

**Introduction**

The interest to magnetoelectric borates has raised recently after the discovery of magnetic field-induced electric polarization in rare-earth (R) ferro-borates RFe$_3$(BO$_3$)$_4$ [1,2] and of strong quadratic magnetoelectric effect in alumo-borates RAl$_3$(BO$_3$)$_4$ [3,4]. Rare earth borates [1,2] thus represent a new class of multiferroic materials [5,6] with non-centrosymmetric crystal structure. These compounds have trigonal structure (space group R32 or P3$_1$21) of natural mineral huntite [7] and consist of helicoidal chains of edge-sharing FeO$_6$ octahera along the trigonal c-axis, interconnected by BO$_3$ triangles and RO$_6$ prisms (upper panel of Fig. 1). The R$^{3+}$ ions occupy D$_3$ symmetry positions inside the isolated prisms and have no direct R-O-R bonds.  Rare-earth borates exhibit several interesting properties strongly depending on the ground state of rare-earth ions in a crystal field [8-13].

The magnetic and magnetoelectric properties of ferroborates RFe$_3$(BO$_3$)$_4$ are rather complex because of the antiferromagnetic ordering of the Fe spins below T$_N$ ~ 30-35 K and their strong coupling with rare-earth moments [8, 9, 14-16]. Besides spontaneous magnetic phase transitions from easy-plane to easy axis states (R=Gd, Ho) [17, 18], they exhibit significant spontaneous magnetic field-induced electric polarization as well as colossal ($\Delta\varepsilon/\varepsilon$ ~ 300%) changes of the dielectric constants below $T_N$ in ferroborates with R = Sm, Ho [19, 20]. As has been shown recently [21], strong magnetoelectric excitation at microwave frequencies (electromagnon) representing one of the antiferromagnetic resonance mode in Fe-subsystem is responsible for this magneto-dielectric effects.

In spite of the significant role of the antiferro-magnetically ordered Fe-subsystem in magnetic and magnetoelectric phenomena of ferroborates, it is the rare-earth subsystem which is responsible for the magnetoelectric coupling, since in $YFe_3(BO_3)_4$ ferroborate with non-magnetic $Y^{3+}$ ions the magnetoelectric properties are very weak [1]. Moreover, the Fe-subsystem in borates has been shown as not essential to establish a large electric polarization in external magnetic fields. Thus, the largest bilinear (quadratic) magnetoelectric effect was observed recently in alumoborates $TmAl_3(BO_3)_4$ [4], $HoAl_3(BO_3)_4$ [3] and $TbAl_3(BO_3)_4$ [22]. The absence of Fe ions and of R-Fe exchange interactions makes $RAl_3(BO_3)_4$ compounds to an appropriate model system to study the rare-earth contribution to magnetoelectric properties. The quantum theory of magnetoelectricity in rare-earth borates was developed in Ref. [23]. This theory is based on a single ion mechanism due to electric dipole moment of rare-earth ion $4f$ shells and due to ion displacement induced by exchange magnetic fields. This approach has been applied to describe temperature and field dependences of polarization in Nd, Sm and Eu ferroborates [23] as well as in alumoborates with R = Tm, Ho, Tb [22, 24]. An important role of low-energy levels, i.e. the ground state of the R-ions in a crystal field of $D_3$ local symmetry has been recognized.

Usually, crystal field (CF) states are determined from the analysis of optical transitions from the ground to exited multiplets [10, 11]. Direct determination of the actual CF levels responsible for the magnetic and magnetoelectric properties requires terahertz (or sub-terahertz) spectroscopic experiments. In addition, magnetic and magnetoelectric properties in real borates crystals may be influenced by local distortions of R-sites due to defects (see for example Ref. [25,26]).

In this work we present the results of terahertz experiments of the low-energy crystal-field states of $Tm^{3+}$ ions in thulium alumo-borate which are responsible for magnetic and magnetoelectric behavior in this material. Magnetic susceptibility of $TmAl_3(BO_3)_4$ reveal strong anisotropy at low temperatures with a significant increase in the basal $ab$-plane (easy directions) compared to the $c$-axis [6, 24, 27]. The observed simultaneous increase of field-induced electric polarization in $TmAl_3(BO_3)_4$ [4] indicates the prevailing influence of the low-energy crystal-field $Tm^{3+}$ states not only on magnetic but also on magnetoelectric properties of the crystal.

Electronic structure of $Tm^{3+}$ in lightly doped $(Y_{1-x}Tm_x)Al_3(BO_3)_4$ crystals has been studied recently by optical absorption and emission spectroscopy [25]. The CF states of the non-Kramers $Tm^{3+}$ ions occupying the positions of $D_3$ symmetry form the singlets $A_{1,2}$ and the doublets E. It was found that the ground CF state of the $Tm^{3+}$ ions is a singlet $A_1$, the next excited state is a doublet (E) at ~29 $cm^{-1}$, and the remaining crystal-field states lie above 100 $cm^{-1}$. Similar structure of the crystal-field states of the ground multiplet $^3H_6$ of $Tm^{3+}$ was also observed in pure $TmAl_3(BO_3)_4$ by optical spectroscopy [27]. Therefore, direct $A_1 \leftrightarrow E$ electronic transitions in the ground multiplet are in the frequency range of the terahertz spectroscopy. The actual energy structure of the low-lying CF $Tm^{3+}$ states in $TmAl_3(BO_3)_4$ was not studied yet. Here we show that the energy levels are affected by local distortions of $Tm^{3+}$ sites in a real crystal and discuss theirs influence on magnetic and quadratic magnetoelectric properties.

**Experiment**

Most results in this work have been obtained using quasi-optical terahertz BWO-spectroscopy [28] in transmittance geometry. The beam path of the terahertz spectrometer is shown in the lower panel of Fig. 1. This spectrometer utilizes linearly polarized monochromatic radiation in the frequency range from 8 $cm^{-1}$ to 37 $cm^{-1}$ which is provided by backward-wave-oscillators (BWO's). BWO's are tunable sources of the electromagnetic radiation based on the periodic modulation of the electron beam in a high-vacuum tube. The variation of the output frequency is realized by changing the accelerating voltage in the range of few kilovolts. The output power of the BWO's is typically 1-100 mW. Opto-acoustic Golay cell or He-cooled bolometer are used as detectors of the radiation.

Frequency dependent transmission spectra have been be analyzed using the Fresnel optical formulas for the transmittance T = |t|$^2$ of a plane parallel sample

$$t = \frac{(1-r^2)t_1}{1-r^2 t_1^2}, \tag{1}$$

where $r = \dfrac{\sqrt{\varepsilon/\mu}-1}{\sqrt{\varepsilon/\mu}+1}$ and $t_1 = \exp\{-2\pi i \sqrt{\varepsilon\mu} d/\lambda\}$.

Here r is the reflection amplitude at the air-sample interface, $t_1$ is the "pure" transmission amplitude, $\varepsilon$ and $\mu$ are the complex dielectric permittivity and magnetic permeability of the sample, respectively, $d$ is the sample thickness, and $\lambda$ is the radiation wavelength. Equation (1) can be applied for anisotropic crystals as well, provided that the incident radiation is linearly polarized along the principal optical axes.

Equation (1) contains two unknown material constants, $\varepsilon$ and $\mu$, and thus cannot be inverted in general case. Fortunately, in present material system the dielectric contribution $\varepsilon$ is only weakly frequency dependent and the magnetic permeability $\mu$ is limited to a narrow frequency range which can be approximated by a sum of Lorentzians:

$$\mu(\nu) = 1 + \Sigma_k \Delta\mu_k \nu_k^2/(\nu_k^2 - \nu^2 + i\nu \Delta\nu_k), \qquad (2)$$

where $\nu_k$, $\Delta\nu_k$ and $\Delta\mu_k$ are resonant frequency, linewidth and the contributions of the $k$-th mode to the magnetic permeability $\mu$.

Direct fitting of the spectra for samples with different thickness using Eqs. (1,2) allows to separate electric and magnetic contributions without additional measurements. In case of transparent samples Fabry-Pérot oscillation on the sample surfaces are visible (see, e.g. Fig. 3 below). The periodicity and amplitude if these oscillations allows to obtain the dielectric permittivity of the sample outside the magnetic modes via Eq. (1).

Static magnetic susceptibility has been measured using commercial Physical Property Measurement System. Single crystals of TmAl$_3$(BO$_3$)$_4$ were grown from flux as described in Ref. [29]. Oriented samples were prepared in the form of plane-parallel *bc*-plane and *ab*-plane oriented plates with several thicknesses from 0.18 up to 1.7 mm and with lateral dimensions of about 10x10 mm$^2$. Transmission experiments on samples with different thickness are necessary to increase the dynamic range of the spectrometer in case of strong absorption. To obtain the mode parameters, the spectra of samples with different thickness have been fitted self-consistently.

**Results and discussion**

Figure 2 shows the static magnetic susceptibility of TmAl$_3$(BO$_3$)$_4$ along the a-axis and along the c-axis. The susceptibility along the a-axis agrees well with Ref. [4, 24, 27] both qualitatively and quantitatively. We note that the c-axis susceptibility in Ref. [4] is roughly by factor of two larger than that shown in Fig. 2, and that measured in Refs. [24, 27], which might be due to small fraction from the a-axis signal. The same effect probably explains small upturn of $\chi_c$ at the lowest temperatures in our data (inset to Fig. 2).

Typical transmittance spectra of TmAl$_3$(BO$_3$)$_4$ in a broad frequency range and in different geometries are demonstrated in Fig. 3. Due to the equivalence of the $\mu$ (and $\varepsilon$) components along the a- and b-axes in the R23 symmetry the geometries shown in Fig. 3 represent full set of possible experimental configurations. Intensive resonance absorption modes are observed around 29 cm$^{-1}$ for ac magnetic field *h* perpendicular to the trigonal c-axis and independently on the electric polarization *e.* This observation suggests the predominant magnetic origin of the excitations. The single mode at 29 cm$^{-1}$ at high temperatures splits into four well-defined components for temperatures below $T$ = 40 K (as denoted in Fig. 4).

From the analysis of the excitations conditions in Figs. 3, 4 we conclude that the observed modes are of purely magnetic origin and no electric components can be detected within the experimental accuracy. However, we cannot exclude that electric contribution will be observed in experiments with higher sensitivity. We estimate possible electric contribution as being at least 15 times smaller than the magnetic one. The parameters of magnetic modes, as obtained by fitting the transmittance spectra of samples with different thickness via Eqs. (1,2), are shown in Fig. 5. The sum of contributions from the magnetic modes in Fig. 5 agrees well with the static magnetic susceptibility within the ab-plane as shown in Fig. 2. Therefore, we conclude that in the influence of the higher-frequency modes on the static properties can in good approximation be neglected. We suggest that present excitations should be attributed to magnetic dipolar CF transitions from the ground singlet A to the next exited doublet E within the ground multiplet $^3H_6$ of Tm$^{3+}$ ions. Their excitation conditions (*h*⊥*c*) correspond to the selection rules (Tab. 1) for transitions A↔E in the CF of rare-earth ion positions with D$_3$ symmetry [30, 31]. As discussed above [25], the transition A$_1$→ E corresponds to the frequency range of the present experiment, the transition energies of other modes exceed 100 cm$^{-1}$.

Table 1. Selection rules for magnetic ($\mu$) and electric (*d*) dipole transitions between crystal field states of non-Kramers rare-earth ion ($Tm^{3+}$) in local sites of $D_3$ symmetry, where $A_{1,2}$ and E are singlets and doublets, respectively [30, 31].

|   | E | $A_1$ | $A_2$ |
|---|---|---|---|
| E | $d_{x,y,z}$, $\mu_{x,y,z}$ | $d_{x,y}$, $\mu_{x,y}$ | $d_{x,y}$, $\mu_{x,y}$ |
| $A_1$ | $d_{x,y}$, $\mu_{x,y}$ | – | $d_z$, $\mu_z$ |
| $A_2$ | $d_{x,y}$, $\mu_{x,y}$ | $d_z$, $\mu_z$ | – |

The contribution to the magnetic permeability of the two components of the $A_1 \leftrightarrow E$ transitions (i.e. $A_1 \leftrightarrow E^+$, $A_1 \leftrightarrow E^-$) can be calculated as [32]:

$$\Delta \mu_{a,b}^{\pm} = 4\pi \cdot N \cdot \frac{2|\mu_{x,y}|^2}{Z(E_1 - E_0)} \left( e^{-\frac{E_1}{k_B T}} - e^{-\frac{E_0}{k_B T}} \right). \quad (3)$$

Here $\mu_x = \mu_y$ are the matrix elements of the $A_1 \leftrightarrow E$ transition for *ac* magnetic field along *x(a)* or *y(b)* axes, respectively; $E_0$ is the ground state energy, $E_1$ is the energy of the doublet; *N* is the concentration of the R-ions, and *Z* is the partition function.

Local distortions of the crystal field of the $Tb^{3+}$ ion are illustrated in Fig. 6. Based on this scheme we attribute the observed structure of the magnetic transitions to local distortions of the crystal field that splits the doublet E into two singlet components $E^{\pm}$. We identify four resonance modes (Figs. 4, 5) as two pairs of transitions $A_1 \rightarrow E_k^{\pm}$ corresponding to two kinds of distorted R-sites (k=1, 2, left panel of Fig. 6). Strongly split components $v_2^{\pm}$ can be assigned to the $Tm^{3+}$ ions being nearest to Bi impurities (right panel of Fig. 6) which arise during the crystal growth and which occupy the R-sites [26,33,34]. The weakly split central components $v_1^{\pm}$ can be attributed to remaining $Tm^{3+}$ ions which are farther away from Bi-impurities and which are subjected to weaker CF distortions. As no further splitting of the spectra is observed, this scheme reasonably explains the sequence of the magnetic transitions in $TmAl_3(BO_4)_3$. Similar $Tm^{3+}$ splitting due to low-symmetry components of the crystal field induced by random lattice strains was observed by optical spectroscopy and analyzed theoretically in Ref. [35]. The line shape in this case was predicted to be asymmetrical due to distribution of distortions. However, in a reasonable approximation two central components can still be described as Lorentzians, Eq. (2), since the mode splitting does not allow to specify the line shape.

In spite of local distortions, the excitation conditions of the observed resonances appear to be similar to the $D_3$ symmetry state. This result is surprising on the first glance. However, it can be explained by averaging of the magnetic response over several equivalent $Tm^{3+}$ sites, which results in the recovery of the initial excitation conditions for undistorted CF states. In contrast to the undistorted sites, the CF transitions at the distorted positions occur between the singlet states A→$E_k^{\pm}$, where $E_k^{\pm}$ are the components of initial doublets ($k$ = 1, 2, see Fig. 6). Thus, the tensor of the complex magnetic permeability $\hat{\mu}^*(\nu) = 1 + \sum \langle \Delta\hat{\mu}_k \rangle \nu_k^2 / (\nu_k^2 - \nu^2 + i\nu\Delta\nu_k)$ in a given frequency range contains four resonance terms at frequencies $\nu_k = \nu_{1\pm}$, $\nu_{2\pm}$ corresponding to the transitions A→$E_1^{\pm}$ and A→$E_2^{\pm}$ at sites #1 and #2, respectively. Here $\langle \Delta\hat{\mu}_k \rangle$ are the contributions of the transitions averaged over equivalent positions with different orientation of the local axes (distortions).

For strongly distorted Tm positions nearest to the Bi impurity (site #2, right panel of Fig. 6), all components of the magnetic tensor $\Delta\hat{\mu}_k$ are nonzero. Here, six $Tm^{3+}$ positions are possible, marked as 1-6 in the right panel of Fig. 6. For $Tm^{3+}$ in the position Nr. 1 the magnetic components in a most general form may be written as:

$$\Delta\hat{\mu} = \begin{pmatrix} \Delta\mu_{xx} & \Delta\mu_{xy} & \Delta\mu_{xz} \\ \Delta\mu_{xy} & \Delta\mu_{yy} & \Delta\mu_{yz} \\ \Delta\mu_{xz} & \Delta\mu_{yz} & \Delta\mu_{zz} \end{pmatrix}, \quad (4)$$

where $x$, $y$, $z$ axes coincide with the crystallographic $a$, $b$, $c$ axes for one of the three equivalent trigonal $a$-axis directions. We omit the transitions indexes ($\pm$) and the site numbers (#1, #2) for simplicity. Due to local $D_3$ symmetry of Bi impurities, other Tm positions shown in Fig. 7 can be obtained starting from the positon Nr. 1 by rotations corresponding to this symmetry. Six symmetry operations, which are necessary in this case, may be described in a unified manner by following transformation matrices:

$$\hat{A}_i = \begin{pmatrix} \cos\alpha_i & \eta_i \sin\alpha_i & 0 \\ -\sin\alpha_i & \eta_i \cos\alpha_i & 0 \\ 0 & 0 & \eta_i \end{pmatrix}, \quad (5)$$

where $\alpha_{1,2,3,4,5,6}$ = 0, 120°, -120°, 0, 120°, –120° are possible rotation angles and $\eta_{1,2,3}=1$, $\eta_{4,5,6}=-1$, respectively.

By adding the contributions from all possible six positions, the average tensor of the magnetic contributions for six equivalent sites #2 from the first coordinate sphere is obtained as:

$$\Delta\hat{\mu}_{(2)} \equiv \langle\Delta\hat{\mu}_{(2)}\rangle = \frac{1}{6}\sum_{i=1...6}\hat{A}_i \cdot \Delta\hat{\mu} \cdot \hat{A}_i^{-1} = \begin{pmatrix} (\Delta\mu_{xx}+\Delta\mu_{yy})/2 & 0 & 0 \\ 0 & (\Delta\mu_{xx}+\Delta\mu_{yy})/2 & 0 \\ 0 & 0 & \Delta\mu_{zz} \end{pmatrix}. \qquad (6)$$

From Eq. (6) we see that the off-diagonal components of general magnetic susceptibility are cancelled out by symmetry and the resulting tensor corresponds to that of $D_3$ symmetry. Therefore, due to averaging over the possible sites the excitation conditions of the magnetic transitions are unaffected by the distortions. In addition, we note that for random distortions closely similar averaging can be performed using continuous distribution.

Using the procedure analogous to that of Eq. (3) the magnetic susceptibility perpendicular to the *c*-axis can be easily obtained. Within the present assumptions the final expression is given by:

$$\Delta\mu_k^{\pm} = 4\pi \cdot c_k N \cdot \frac{2|\mu_k^{\pm}|^2}{Z(E_k^{\pm}-E_0)}\left(e^{-\frac{E_0}{k_BT}} - e^{-\frac{E_k^{\pm}}{k_BT}}\right), \qquad (7)$$

where $|\mu_k^{\pm}|^2 = (|\mu_{xk}^{\pm}|^2 + |\mu_{yk}^{\pm}|^2)/2$ are effective matrix elements of the magnetic dipolar transitions averaged over six equivalent $Tm^{3+}$ sites, $E_0$ and $E_k^{\pm}$ are the CF states of the ground singlet and the two components of the split doublet, respectively, $c_k$ is the concentration of the rare-earth ions of distorted sites of *k*-type. Equation (7) includes contributions of the CF transitions from the ground singlet to the doublet components $E_i^{\pm}$. It replaces Eq. (3) in case where distortions must be included into the consideration. The magnetic contributions $\Delta\mu_k^{\pm}$ of the observed modes is shown by solid lines in the lower panel of Fig. 5. The calculations are shown by solid lines and they reproduce well the temperature dependence of the experimental data. The effective matrix elements in a first approximations are $|\mu_1^{\pm}| \approx 3.6\mu_B$, $|\mu_2^{+}| \approx 4.4\mu_B$, $|\mu_2^{-}| \approx 2.8\mu_B$, and the corresponding concentrations of the distorted $Tm^{3+}$ ions are obtained as $c_1 = 1 - 6x$, $c_2 = 6x$, where $x = 3.6\%$ is the Bi concentration. This value of x agrees well with the chemical composition of the crystal as obtained by the X-ray energy dispersion analysis giving Bi concentration $x \approx (5\pm2)\%$.

Within a good approximation, the splitting of the magnetic modes does not influence the static magnetic susceptibility which is simply given by the sum of all observed contributions. The calculated total contribution of four A-$E_k^{\pm}$ transitions to the static susceptibility along the *a*-axis is given by dashed line in Fig. 2 and is provides good description of the experimental data. In present experiments no dynamic magnetic contribution along the c-axis has been observed within the sensitivity of our spectrometer. This agrees with the weakness of the static susceptibility along the c-axis, $\chi_c \ll \chi_a$, as shown in Fig. 2. The remaining value of $\chi_c$ can be explained by quadratic terms in $\Delta\mu_z$ as a function of crystal-field perturbations, while the observed splitting of the doublet states E is linear with respect to these perturbations.

The splitting of the crystal-field states of $Tm^{3+}$ ions was confirmed by experiments in external magnetic fields. Representative spectra in external magnetic fields parallel to the c-axis are shown in Fig. 8. External magnetic field additionally splits the $E_i^{\pm}$ singlets as:

$$E_i^{\pm}(H_c) = E_i^{-} + \Delta/2 \pm \sqrt{\Delta^2 + \mu_c^2 H_c^2} \ , \qquad (8)$$

where $\Delta = E_i^{+} - E_i^{-}$ is the doublet splitting due to the distortions of the local CF (Fig. 6), $\mu_c$ is the diagonal matrix element of the undisturbed doublet states. Fitting the field dependencies of the resonance frequencies $\nu^{\pm} = E^{\pm}/(h \cdot c)$ (Fig. 9) to Eq. (8) gives the matrix element along the c-axis as $\mu_c$ = (0.9 ± 0.05)$\mu_B$. This value of $\mu_c$ is in a good agreement with that calculated from the CF parameters [24] and it allows to evaluate the contribution to the magnetic properties along the c-axis from the doublet states. The calculation results are given by the dashed line in the inset to Fig. 2. As the theoretical values are clearly below the experimentally obtained $\chi_c$, the contribution from higher frequency transitions are relevant in this case.

Finally, it is instructive to analyze the effect of the observed distortions of local $Tm^{3+}$ sites on the magnetoelectric properties of the crystal. Within the R32 space group of an undistorted crystal and according to Ref. [15], the magnetic field-induced electric polarization of alumo-borates in the ab-plane is determined by the expressions:

$$\begin{aligned} P_x &= \lambda_1 H_y H_z + \lambda_2 (H_x^2 - H_y^2)... \\ P_y &= -\lambda_1 H_x H_z - 2\lambda_2 H_x H_y ... \end{aligned} \qquad (9)$$

Here "…" replaces the omitted higher-order terms. For moderate fields $\lambda_1$ and $\lambda_2$ can be considered as quadratic magnetoelectric susceptibilities. However, in a distorted crystal the local magnetoelectric response has to include further components:

$$\begin{aligned} P_x^{loc} &= \lambda_{xxx} H_x^2 + \lambda_{xyy} H_y^2 + \lambda_{xzz} H_z^2 + \lambda_{xxy} H_x H_y + \lambda_{xxy} H_x H_y + \lambda_{xxz} H_x H_z + \lambda_{xyz} H_y H_z ... \\ P_y^{loc} &= \lambda_{yxx} H_x^2 + \lambda_{yyy} H_y^2 + \lambda_{yzz} H_z^2 + \lambda_{yxy} H_x H_y + \lambda_{yxy} H_x H_y + \lambda_{yxz} H_x H_z + \lambda_{yyz} H_y H_z ... \end{aligned}, \qquad (10)$$

where $\lambda_{xxx}, \lambda_{xyy}$, etc. are local magnetoelectric susceptibilities for a given type of a distorted $Tm^{3+}$ site with magnetic field oriented with respect to local axes of corresponding equivalent positions.

Here the position indexes are omitted for clarity. Total macroscopic polarization is obtained by averaging over all positions and taking into account the interrelation of local axes as determined by Eq. (5). After the averaging procedure, Eq. (10) has the form of Eq. (9) with effective magnetoelectric susceptibilities given by:

$$\lambda_1^{eff} = (\lambda_{xyz} - \lambda_{yxz})/2, \text{ and } \lambda_2^{eff} = (\lambda_{xxx} - \lambda_{xyy} - \lambda_{yxy})/4 \qquad (11)$$

Therefore, in spite of local distortions of $Tm^{3+}$ sites, the symmetry of the macroscopic magnetoelectric response is recovered to R32 symmetry with renormalized magnetoelectric susceptibilities. In the absence of distortions, the corresponding local magnetoelectric susceptibilities in Eq. (11) are reduced to $\lambda_1$ and $\lambda_2$ with $\lambda_{xyz} = -\lambda_{yxz} \equiv \lambda_1$ and $\lambda_{xxx} = -\lambda_{xyy} = -\lambda_{yxy}/2 \equiv \lambda_2$, respectively. Eq. (9) reproduces the quadratic behavior of the electric polarization along the a-axis in external magnetic field $H \parallel a$ as observed in Ref. [4]. In experiments of Ref. [4] similar field dependence has been detected for $P \parallel c$ clearly contradicting the predictions by Eq. (9). Further investigations are necessary to resolve this problem.

**Conclusions**

Crystal field (CF) transitions of the ground multiplet $^3H_6$ of $Tm^{3+}$ ions have been investigated in the magnetoelectric $TmAl_3(BO_3)_4$ and in the terahertz frequency range. Pure magnetic character of the observed modes has been confirmed. Fine structure of the transitions has been detected and assigned to distortions of the local CF by $Bi^{3+}$ impurities. We suggest a model description of the distorted state which explains quantitatively the temperature and magnetic field dependencies of the mode parameters. The values of the matrix elements of the magnetic transitions are obtained and the contributions of magnetic dipolar transitions to the static magnetic properties are analyzed. Due to site averaging, even in the presence of distortions the symmetry of the magnetic response as well as quadratic magnetoelectric susceptibility correspond to that of an undistorted crystal. We believe that the recovery of macroscopic symmetry of physical properties in crystals with random local distortions could also be observed in other systems.

**Acknowledgements**

This work was partially supported by the Russian Foundation for BasicResearch (14-02-91000, 15-02-07647, 14-02-00307, 13-02-12442, 15-42-04186), Sci. school-924.2014.2, and by the Austrian Science Funds (I815-N16, W-1243, P27098-N27).

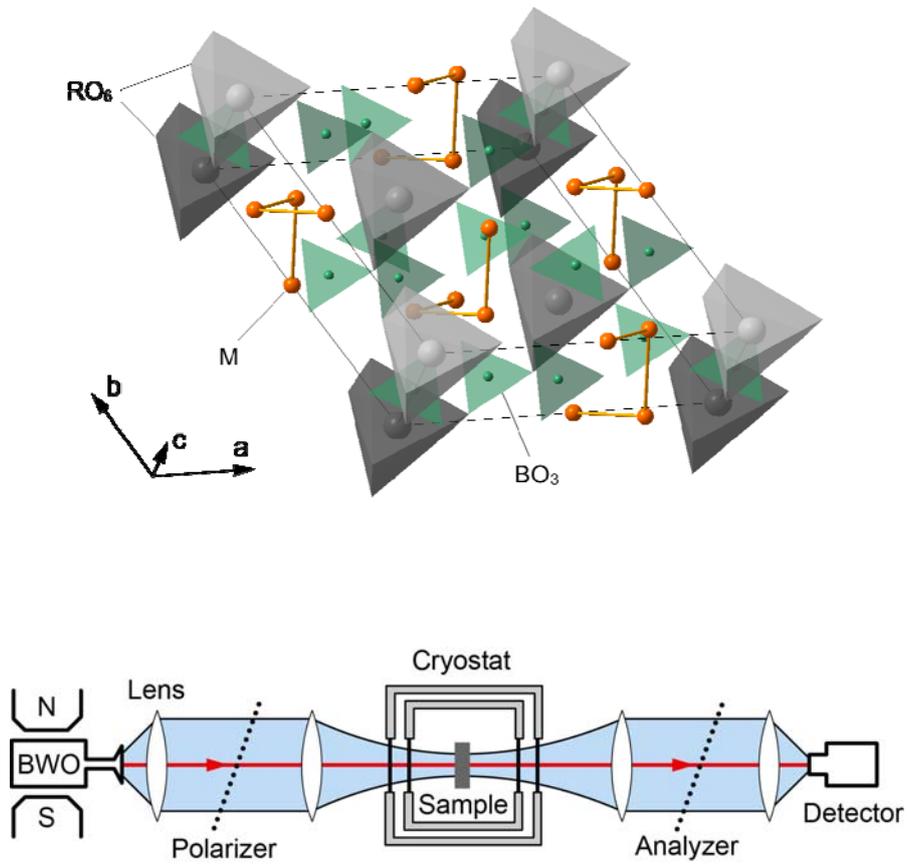

Fig. 1. Upper panel: crystal structure of rare earth borates $RM_3(BO_3)_4$ (R = Y, La, Pr-Yb; M = Fe, Al). Lower panel: scheme of the quasi-optical transmission spectrometer on the basis of backward wave oscillator (BWO) utilized in this work.

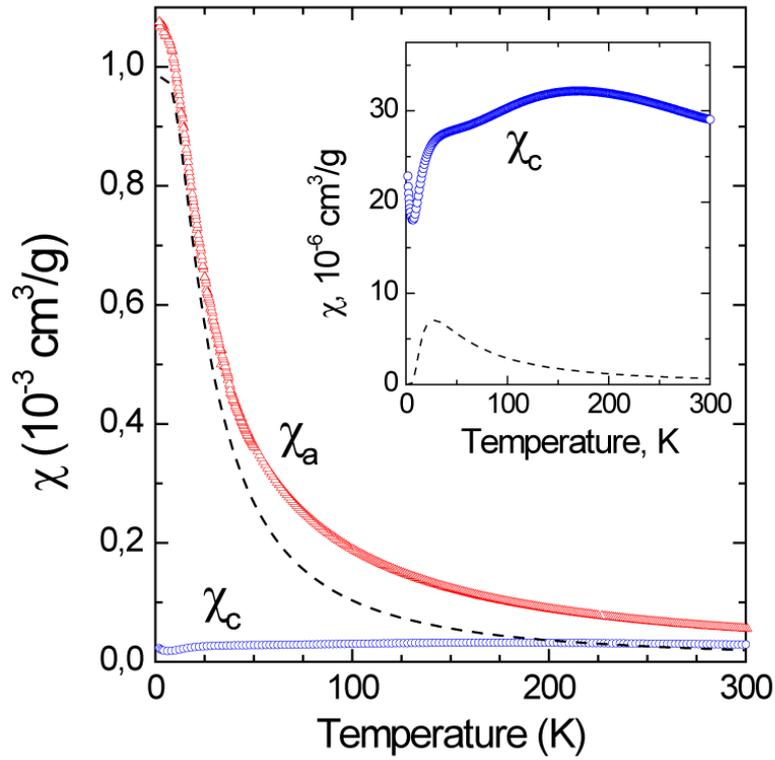

Fig. 2. Static magnetic susceptibility of TmAl$_3$(BO$_3$)$_4$ along the *a* and *c* axes measured in magnetic field $\mu_0 H$ = 0.1 T. Symbols – experiment, dashed line shows the calculated contributions from the observed A-E transitions as described in the text, Eq. (7). The inset shows an expanded view of the *c*-axis susceptibility. The dashed line represents a contribution from the transitions of the first excited Tm$^{3+}$ doublet which becomes frozen below 25 K explaining rapid decrease of $\chi_c$ at low temperatures.

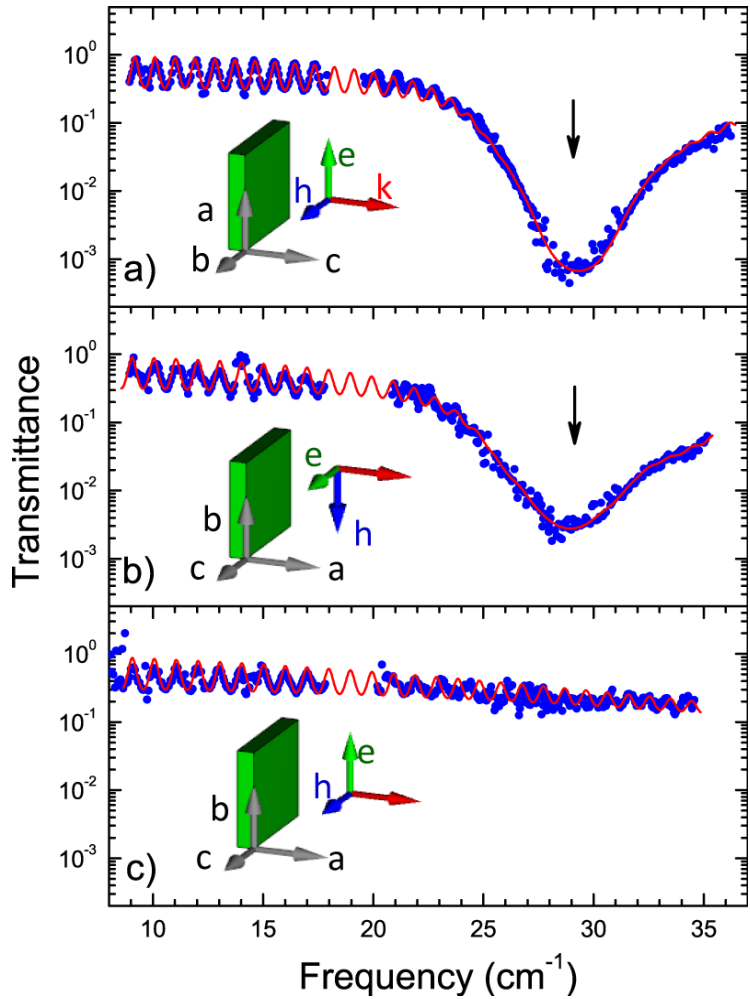

Fig. 3. Terahertz transmittance spectra of *ab*-plane (a) and *bc*-plane (b, c) oriented TmAl$_3$(BO$_3$)$_4$ for different orientations of the *ac* **h** and **e** fields with respect to the crystallographic axes and at *T* = 60 K. Points – experiment, solid lines are obtained via Fresnel equations with magnetic contribution given by Eq. (2). Sample thickness: *ab*-cut: 1.71 mm, *bc*-cut: 1.54 mm. The oscillations in the spectra are due to the Fabry-Pérot interferences on the sample surfaces.

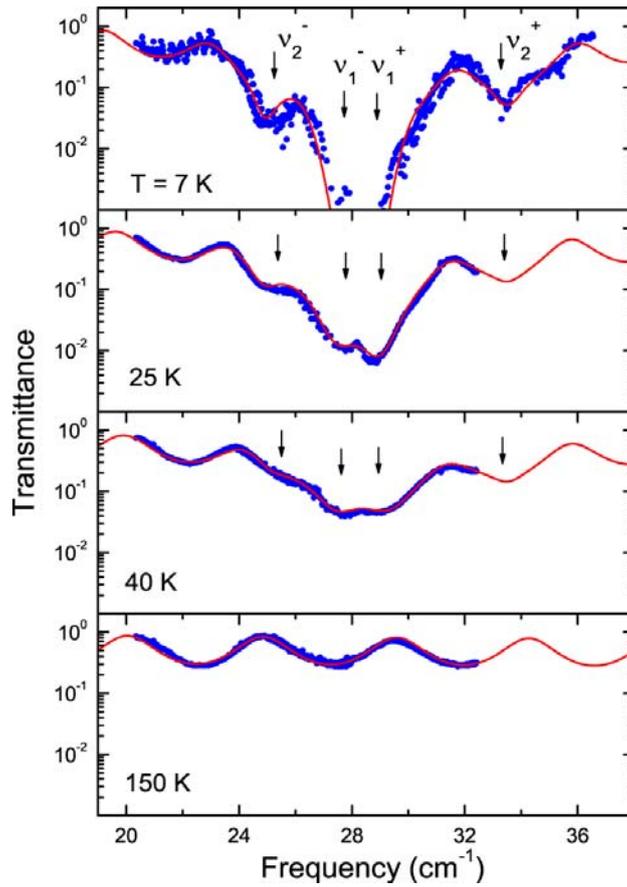

Fig. 4. Transmission spectra of thin *bc*-oriented plate of TmAl$_3$(BO$_3$)$_4$ for *ac* magnetic field **h**⊥*c*-axis demonstrating the emerging splitting of the magnetic transition. Points – experiment, solid lines are obtained via Fresnel equations, Eq. (1) with magnetic contribution given by Eq. (2). Smaller periodicity of the Fabry-Pérot oscillations is due to different sample thickness (d = 0.31 mm) compared to Fig. 3.

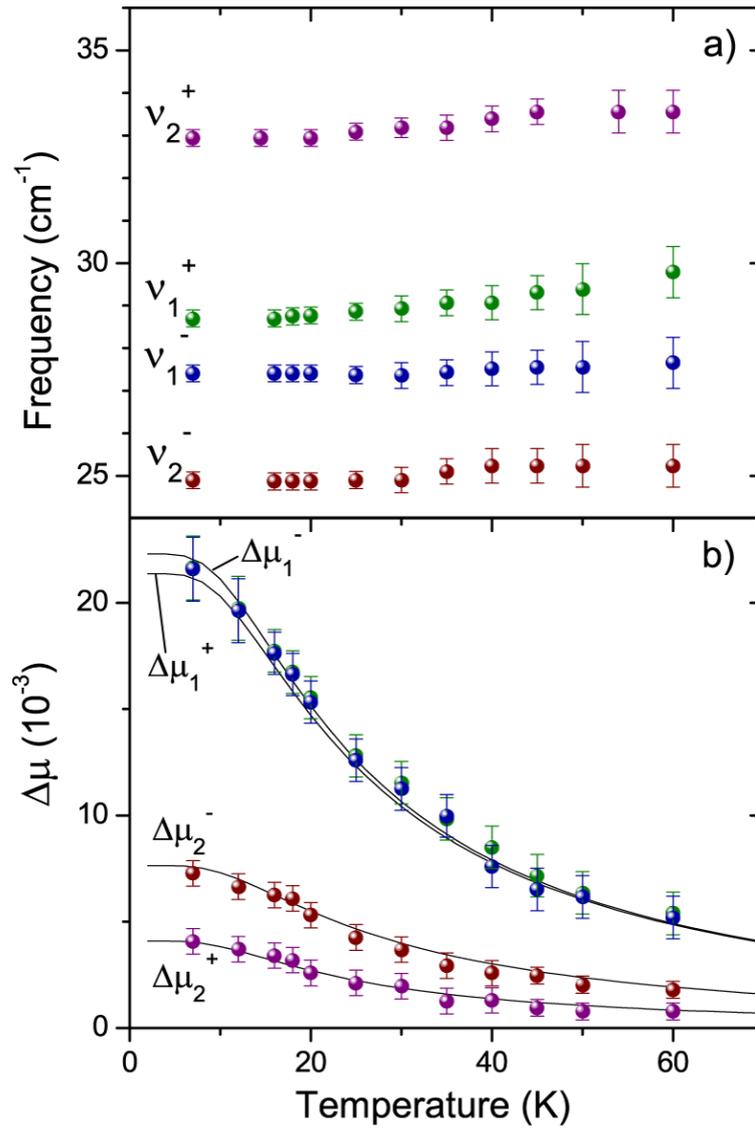

Fig. 5. Temperature dependencies of the resonant frequencies in TmAl$_3$(BO$_3$)$_4$ (a) and of the contributions to the magnetic permeability (b) of the resonance modes observed in Fig. 4. Symbols represent the experimental values obtained from the terahertz transmission spectra, solid lines in (b) are calculated according to Eq. (7).

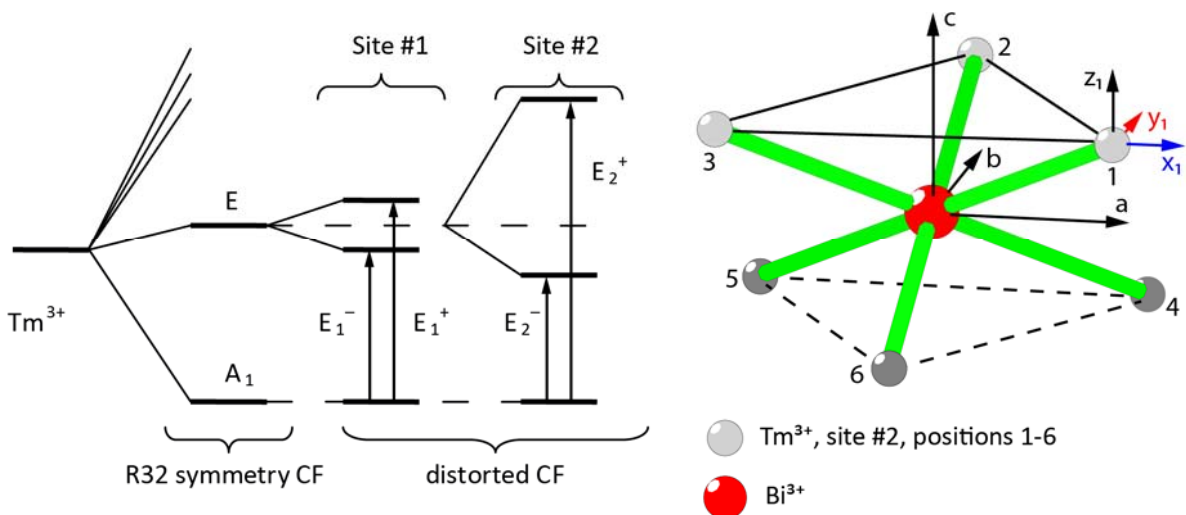

Fig. 6. (left) Splitting of the crystal field states of $Tm^{3+}$ ions in the vicinity of a Bi-impurity. (right) Local arrangement of six possible positions of strongly distorted $Tm^{3+}$ ions (site #2) around a single Bi-impurity. $Bi^{3+}$ and $Tm^{3+}$ lie in the middle of $RO_6$ prisms (Fig. 1) which are not shown for simplicity.

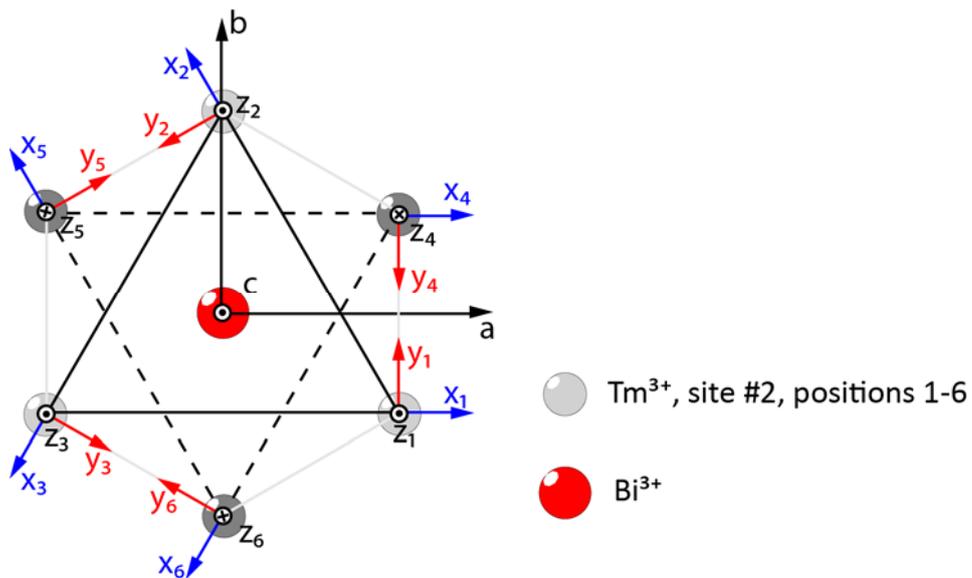

Fig. 7. Projection of six $Tm^{3+}$ positions close to a single Bi-impurity (sites #2, Fig. 6, right panel) to the crystallographic ab-plane and the orientation of the local axes $r_i$ of corresponding positions with respect to crystallographic axes $r$. Symmetry transformation between the six positions are determined by the rotation matrixes $\hat{A}_i$ (i=1…6), Eq. (5): $\vec{r}_i = \hat{A}_i \vec{r}$ as described in the text.

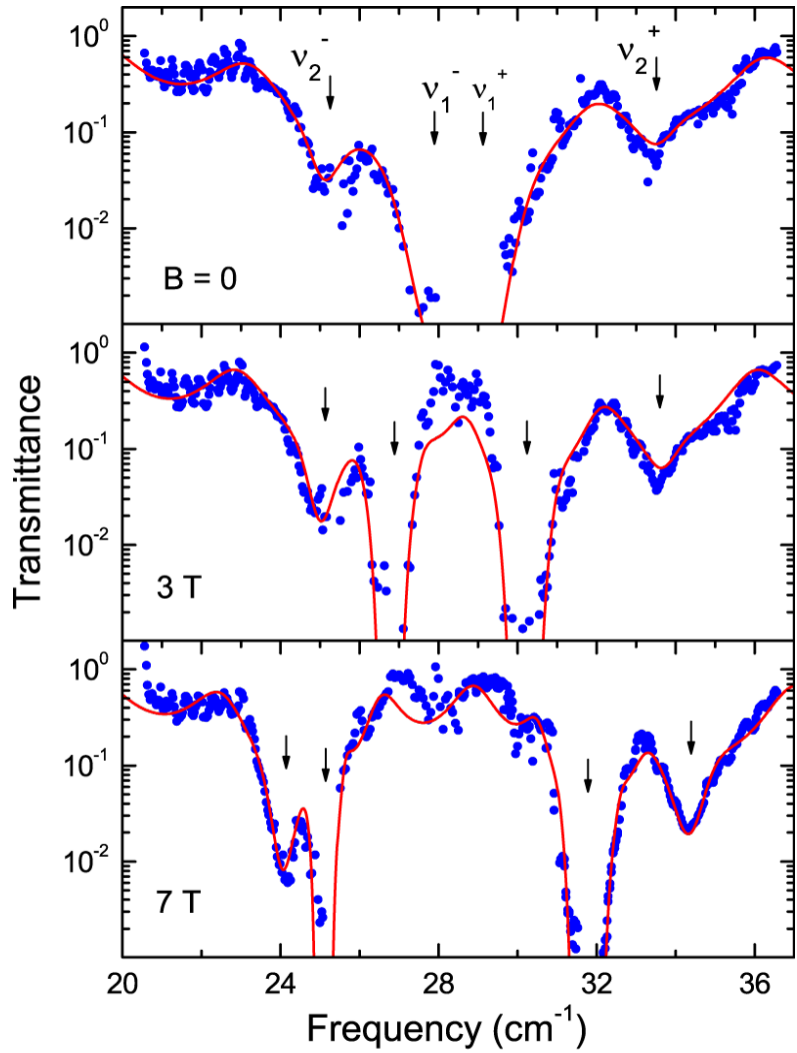

Fig. 8. Transmittance spectra of TmAl$_3$(BO$_3$)$_4$ measured in static magnetic fields parallel to the *c*-axis and for *ac* magnetic field **h**⊥*c*-axis. *T* = 7 K, *d* = 0.31 mm. Points – experiment, solid lines are obtained via Fresnel equations, Eq. (1), with magnetic contribution given by Eq. (2). Arrows indicate the positions of the resonance modes.

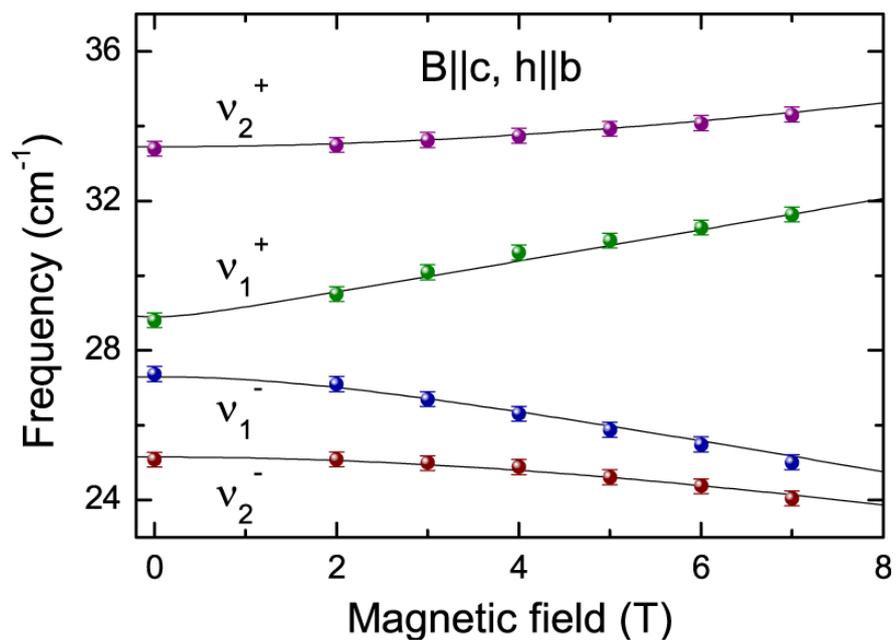

Fig. 9. Magnetic field dependence of the resonance frequencies in TmAl$_3$(BO$_3$)$_4$. Symbols – experimental data as obtained from the fit parameters in Fig. 8, lines – theory according to Eq. (8).